\documentclass[10pt]{cai}
\usepackage[capitalise, nameinlink]{cleveref}
\usepackage{scrextend}

\def\boxit#1{\vbox{\hrule\hbox{\vrule\kern6pt
          \vbox{\kern6pt#1\kern6pt}\kern6pt\vrule}\hrule}}

\definecolor{orange}{rgb}{1,0.5,0}
\definecolor{MyDarkBlue}{rgb}{0,0.08,0.45}

\begin{document}
\begin{center}

\title{Deep Hedging with Market Impact\footnotemark[1]}
\maketitle

\thispagestyle{empty}

\begin{tabular}{cc}
Andrei Neagu\textsuperscript{2}
\upstairs{\affilone}, Fr\'ed\'eric Godin\upstairs{\affiltwo}, Clarence Simard\upstairs{\affilthree}, Leila Kosseim\upstairs{\affilone}
\\[0.25ex]
{\small \upstairs{\affilone} Concordia University, Department of Computer Science and Software Engineering, Montreal, Canada} \\
{\small \upstairs{\affiltwo} Concordia University, Department of Mathematics \& Statistics, Montreal, Canada} \\
{\small \upstairs{\affilthree} Universit\'e du Qu\'ebec \`{a} Montr\'eal, Département de Mathématiques, Montreal, Canada} \\
\end{tabular}

\deffootnote[2em]{0em}{1em}{\textsuperscript{\thefootnotemark}\,}
\footnotetext[1]{Andrei Neagu is supported by the NSERC Canada Graduate Scholarships-Master’s (CGS M). Godin is supported by NSERC (RGPIN-2017-06837). Simard is supported by NSERC (RGPIN-2020-06619). Kosseim is supported by NSERC (RGPIN-2020-05542). We thank Alexandre Carbonneau for knowledge transfer that helped us throughout the conduction of this research project.}
\footnotetext[2]{ Corresponding author. \vspace{0.2em} \\ \mbox{\hspace{0.47cm}} \textit{Email addresses:} andrei.neagu@concordia.ca (Andrei Neagu), frederic.godin@concordia.ca (Fr\'ed\'eric Godin), simard.clarence@uqam.ca (Clarence Simard), leila.kosseim@concordia.ca (Leila Kosseim).}
\vspace*{0.2in}
\end{center}

\begin{abstract}
Dynamic hedging is the practice of periodically transacting financial instruments to offset the risk caused by an investment or a liability. Dynamic hedging optimization can be framed as a sequential decision problem; thus, Reinforcement Learning (RL) models were recently proposed to tackle this task. However, existing RL works for hedging do not consider market impact caused by the finite liquidity of traded instruments. Integrating such feature can be crucial to achieve optimal performance when hedging options on stocks with limited liquidity. In this paper, we propose a novel general market impact dynamic hedging model based on Deep Reinforcement Learning (DRL) that considers several realistic features such as convex market impacts, and impact persistence through time. The optimal policy obtained from the DRL model is analysed using several option hedging simulations and compared to commonly used procedures such as delta hedging. Results show our DRL model behaves better in contexts of low liquidity by, among others: 1) learning the extent to which portfolio rebalancing actions should be dampened or delayed to avoid high costs, 2) factoring in the impact of features not considered by conventional approaches, such as previous hedging errors through the portfolio value, and the underlying asset's drift (i.e. the magnitude of its expected return).
\end{abstract}


\section{Introduction}

Hedging is a trading strategy that consists of purchasing or selling shares of a financial instrument to offset the risk associated with a correlated financial asset. Dynamic hedging is a strategy where the number of hedging instrument shares in the portfolio is regularly adjusted (rebalanced) through time to improve risk mitigation. Dynamic hedging can be framed as a sequential decision making problem, with actions being the transactions performed on each portfolio rebalancing date, and states representing the current market conditions.

In practice, transactions performed on the market (for hedging or other purposes) adversely impact the asset price. If the number of shares requested for buying (or offered for selling) at a certain price is insufficient to cover the supply (or demand), then orders from other market participants at the next best prices are used to complete the transaction. The larger the transaction is, the furthest away the actual transaction price will be from the original best available price on the market. This phenomenon is due to \textit{finite liquidity}, which describes the scarcity of available orders, and is known as the \textit{market impact}. This market impact can either be immediate, or it can persist over time. The persistence of the market impact over time is known as the \textit{impact persistence}.

In this paper, we address the optimization of the trading strategy of an agent applying a dynamic procedure to hedge a position on a financial asset in the presence of market impacts with different levels of impact persistence. We propose a Deep Reinforcement Learning (DRL) procedure to tackle dynamic hedging optimization. DRL has the ability to work in conjunction with very general market models, while being able to handle the high-dimensional state spaces needed to provide a full depiction of the processes driving market impacts. While dynamic hedging procedures based on reinforcement learning have received attention recently (e.g. \cite{buehler2019deep, carbonneau2021deep, carbonneau2021equal, cao2023gamma,wu2023robust}), realistic models of market impacts have not yet been fully integrated in such frameworks.

The optimal trading policy produced by our proposed DRL approach is compared to more common strategies and hedging baselines, and is shown to better adapt to an environment which includes market impacts. Results from simulations show that the model allows: 1) learning the extent to which portfolio rebalancing actions should be dampened or delayed to avoid high costs, 2) factoring in the impact of features not considered by conventional approaches, such as previous hedging errors through the portfolio value, and the underlying asset's drift (i.e. the magnitude of its expected return).

\section{Background}

\subsection{Previous Work on Hedging and Market Impact}
\label{se:Hedging}

Since the 1970s, the standard approach to dynamic hedging of options has been \textit{delta hedging} \cite{BSM1973, hull1993options}. This strategy consists in setting the number of hedging instrument shares to be included in the hedging portfolio in such a way that the first-order partial derivative of the net position, which includes the liability and the hedging instrument, with respect to the option underlying asset price, is zero. Delta hedging has shown to produce optimal risk mitigation in the idealized setting of \cite{BSM1973}. However, in more realistic market environments taking market frictions (such as market impact) into consideration, delta hedging leads to sub-optimal results because it is myopic and it leads to excessive trading \cite{godin2016minimizing}. 

Several generalizations of the delta hedging approach have been proposed to address these shortcomings. For example, the Leland delta hedge \cite{leland1985} adjusts hedging positions (i.e. the number of hedging instrument shares to hold) to take proportional transaction costs into consideration. On the other hand, the variance-optimal strategy of \cite{schweizer1995variance} minimizes the variance of multi-period hedging losses in absence of market impact. This strand of literature culminated in the Deep Reinforcement Learning (DRL) approach of \cite{buehler2019deep} which is applicable for a collection of risk measures, general dynamics for asset prices and highly multi-dimensional state spaces for the market environment. However, the latter work only considers proportional transaction costs and does not model market impacts stemming from finite liquidity.

Another stream of literature has focused on modelling market impacts of transactions in the context of finite liquidity, such as \cite{Cetin/Jarrow/Protter_2004} who proposed to use supply curves to model market impacts. Dynamic transaction optimization in the presence of market impact models have been investigated in several works, such as \cite{almgren2001optimal, cartea2015optimal, ning2021double} for optimal execution, \cite{Calvez/Cliff_2018} for trading, \cite{Guasoni/Weber_2020} for optimal investing and \cite{Bank/Baum_2004, Cetin/Soner/Touzi_2010, Rogers/Singh_2010,  Almgren/Li_2016, gueant2017option, Ellersgaard/Tegner_2017} 
for hedging optimization. However, most hedging works in this stream of literature rely on continuous-time stochastic control methods (e.g. continuous-time dynamic programming) that require a very deep understanding of complex probabilistic methods and are therefore of limited accessibility for practitioners. Furthermore, obtaining solutions with optimal control techniques often requires making compromises in terms of model realism. The DRL approach proposed in this paper addresses these issues as it can be adapted to work under very general market impact dynamics. 





\subsection{Financial Background} 
\label{se:marketdyn}
This section first provides financial background knowledge and describes the market dynamics considered in our work, which are borrowed from \cite{simard2014general}.
For the reader's convenience, all notations used in this paper are summarized in \cref{tb:notations}.

An option is a financial contract which gives the possibility to purchase (if the option is a \textit{call} option) or to sell (if a \textit{put} option) an asset, referred to as the underlying asset, at a predetermined price (the \textit{strike price} $K$) and a given date $T$ (the \textit{expiry}). The \textit{time-to-maturity}, denoted $T-t$, is the time left before the expiry. The value of the option is related to the underlying asset value, with movements in the price of the latter causing either an increase or a decrease in the option value. Thus, such a correlation can be used by a financial institution which issued an option and for which the option value is a liability. It can transact shares of the underlying asset to offset the risk of a potential appreciation of the option value.

\begin{table}
\begin{tabular}{| c | l || c | l |}
\hline
$\mathbf{t}$ & \textbf{time step} & $\boldsymbol{X}$ & \textbf{hedging position (number of shares)}\\
\hline
$K$ & strike price & $T$ & expiry\\
\hline
$\pi_0$ & option premium & $F^a$ & cost of buying\\
\hline
$F^b$ & revenue from selling & $\mathbf{A,B}$ & \textbf{impact persistence (buy/sell)}\\
\hline
$\rho$ & risk measure & $R$ & random loss variable\\
\hline
$\mathcal{P}_X$ & profits & $M$ & cash reserve\\
\hline
$r$ & risk-free rate & $e^r$ & accrual factor\\
\hline
$c$ & transaction total amount & $\mathbf{S}$ & \textbf{underlying asset price}\\
\hline
$\mu$ & drift (annualized expected return) & $\sigma$ & volatility\\
\hline
$\alpha,\beta$ & market impact parameters (buy/sell) & $\lambda_a,\lambda_b$ & impact persistence parameters (buy/sell)\\
\hline
$\mathbf{V}$ & \textbf{hedging portfolio value} & $\delta_t$ & time step length\\
\hline
\end{tabular}
\caption{Symbols used and their definition, with quantities used as state variables in the DRL model being denoted in bold.}
\label{tb:notations}
\end{table}


In this paper, we consider an agent who sold a call option, for which they received a cash amount, called a \textit{premium} $\pi_0$, and who wishes to hedge the risk of the option value increasing by periodically purchasing or selling shares of the underlying asset. The agent's \textbf{action} consists in adjusting the number of shares it holds at each time step. The set of time steps is denoted by $\mathcal{T}=\{0,\ldots,T\}$. The underlying asset's (a stock) liquidity is limited, with trades of the agent causing market impacts on its price. The \textit{impact persistence} state variable $y$ characterizes the persistence of the impact resulting from the agent's previous transactions on the stock. The cost of purchasing $x$ shares of the stock at time $t$ when the \textit{impact persistence} state variable is $y$ is denoted $F^a(t,y,x)$, while the revenue from selling $x$ shares at time $t$ is given by $F^b(t,y,x)$. The function $F^a(t,y,x)$ (resp. $F^b(t,y,x)$) needs to be increasing and convex (resp. decreasing and concave) with respect to both $x$ and $y$ to properly represent market impacts. Functions $F^a$ and $F^b$ characterize the structure of the limit order book in the model, which consists in the set of numbers of available stock shares for buy and sell orders at each price.
The evolution of the impact persistence variables $y$ on the buy and sell sides are assumed to be driven respectively by the two processes $\{A_t\}_{t \in \mathcal{T}}$ and $\{B_t\}_{t \in \mathcal{T}}$ whose specifications are discussed in \cref{se:pricesAndImpact}.





The dynamic hedge is performed by setting up a portfolio invested in cash and underlying asset shares, and whose final value is meant to offset potential losses on the sold option position at expiry. The number of shares in the portfolio is represented by the process $X=\{X_t\}_{t=1,\ldots,T}$, where $X_t$ corresponds to the number of stock shares in the portfolio during $(t-1,t]$, also known as the \textit{hedging position}.
The dynamic hedging problem consists of selecting a trading strategy $X$ leading to the smallest possible risk at maturity, i.e. solving
\begin{equation}
\label{globhedging}
X^*=\underset{X}{\arg\min} \, \rho \left( -\mathcal{P}_X\right)
\end{equation}
where $\rho$ is a chosen risk measure mapping a random loss variable $R$ into a real number representing perceived risk, and where $R=-\mathcal{P}_X$ is the negative of the total profits for the agent immediately after the maturity of the option if the hedging strategy $X$ is pursued. Several risk measures $\rho$ have been used in the literature \cite{hodges1989optimal, gueant2017option, carbonneau2021deep}. In this work, we consider the \textit{semi-quadratic penalty} risk measure $\rho^{semi}(R)=\mathbb{E}\left[ (R\mathds{1}_{ \{R>0\} }  )^2  \right]$.
The popular quadratic penalty is not considered, since it has the issue of penalizing gains, a problem the semi-quadratic penalty does not have.  This drawback is more problematic in the presence of market impacts, as the optimal strategy tends to artificially increase turnover after making hedging gains to incur market impacts and avoid positive cumulative gains at maturity.

The hedging portfolio is assumed to be self-financing, i.e. all purchases or sales of stock shares are financed through the portfolio cash reserve process $M$. The quantity $M_t$ denotes the amount of cash within the portfolio at time $t$ right before the portfolio is rebalanced.
Between transactions, the cash reserve is set to accrue (increase) at the one-period risk-free rate $r$ (the accrual factor being $e^r$) on each time period.
The cash amount in the portfolio can be found recursively through:
\begin{eqnarray}
M_t = 
\begin{cases}
\pi_0, \quad  \quad\, t=0,
\\ \left(M_{t-1} - c_{t-1}(X)\right)e^r, \quad t=1,\ldots,T,
\end{cases}
\end{eqnarray}
with the time-$t$ stock transaction total amount due to rebalancing being defined as:
\begin{equation}
c_t(X)\equiv
\begin{cases}
F^a(t,A_t,\Delta X^a_{t+1}) - F^b(t,B_t,\Delta X^b_{t+1}), \quad t=0,\ldots,T-1,
\\ 0 \text{ if } t=T,
\end{cases}
\end{equation}
where transactions on the stock are characterized by:
\begin{equation}
\Delta X_t \equiv X_t - X_{t-1}, \quad
 X^a_t \equiv \sum^t_{i=1} (\Delta X_i)^+, \quad
 X^b_t \equiv \sum^t_{i=1} (\Delta X_i)^-,
\end{equation}
with $x^+ \equiv \max(0,x)$, $x^-\equiv \max(0,-x)$ and $X_0 = 0$. The process $\Delta X$ characterizes period-by-period variations in the number of shares within the portfolio and $X^a$ and $X^b$ represent respectively the total number of shares purchased and sold since the initiation of the hedge. Note that $X_t = X^a_t - X^b_t$. 

At the maturity of the hedge $T$, the buyer, who the agent sold the option to, will choose to exercise their right to buy the underlying asset at the strike price $K$ only if the revenue from selling one share of the underlying asset is larger than the strike price $K$ (i.e. $F^b(T,B_T,1)>K$). Thus, we can define the exercise event $E \equiv\{F^b(T,B_T,1)>K\}$. After implementing the hedging strategy $X$, the total profit for the agent right after the option maturity is:
\begin{align}
\begin{split}
\mathcal{P}_X \equiv F^b\left(T,B_T,(X_T - \mathds{1}_{E})^+\right)-F^a\left(T,A_T,(-X_T+\mathds{1}_{E})^+\right) + M_T + K \mathds{1}_{E},
\end{split}
\end{align}
with $\mathds{1}_{E}$ being the dummy indicator worth $1$ if and only if event $E$ occurs. The term containing $F^b$ represents stock sales proceedings if the agent has more than one share in his portfolio after the settlement. Conversely, the $F^a$ term represents the cost of purchasing stocks to cover outstanding sell positions after the settlement.

\vspace{-5pt}
\subsection{Market Environment}
\label{se:pricesAndImpact}

For illustrative purposes and without loss of generality, we consider a more specific model taken from \cite{simard2014general} for price dynamics and market impacts, which acts as our simulated market environment in the numerical examples of the subsequent sections. First, a price process $S \equiv \{S_t\}_{t \in \mathcal{T}}$ is considered, which reflects the price of the underlying asset that would be obtained without any market impact. In order to generate a simulation dataset for evaluation purposes, we used the geometric Brownian motion (GBM) to simulate the underlying asset prices $S_t$ at each time step $t$:
\begin{equation}
\label{Bsmodel}
S_t = S_{t-1} \exp \left(\left( \mu - \sigma^2/2 \right) \delta_t + \sigma \sqrt{\delta_t} Z_t\right)
\end{equation}
where $\{Z_t\}^T_{t=1}$ are independent standard Gaussian random variables, $\mu$ and $\sigma$ are respectively the annualized expected return (also known as the drift) and the volatility of the underlying asset, and $\delta_t$ is the time elapsing (in years) between any two time points $t$.  
Furthermore
\begin{equation}
\label{dynamicsF}
F^u(t,y,x) \equiv G^u(t,x+y)-G^u(t,y), \quad u\in \{a,b\}
\end{equation}
where $G$ is defined as
\begin{equation}
\label{dynamicsG}
G^a(t,x)=S_t\left\{(1+x)^\alpha-1 \right\}, \quad G^b(t,x) \equiv S_t\left\{(1+x)^\beta-1 \right\}
\end{equation}
for some parameters $\alpha>1,\beta\in(0,1)$ reflecting the degree of market illiquidity.
Moreover, the impact persistence processes for buying and selling ($A$ and $B$), which describe the way the impact of a transaction persists over time, are characterized by
\begin{equation}
A_{t}=e^{-\lambda_a} \left(A_{t-1} + \left(\Delta X_t\right)^+\right), \quad B_{t}= e^{-\lambda_b}\left( B_{t-1} + \left(\Delta X_t\right)^-\right), \quad t=1,\ldots,T
\end{equation}
for given values $A_0\geq0$, $B_0 \geq 0$ and some parameters $\lambda_a,\lambda_b>0$ which entail an exponentially decaying impact. A case of interest is $\lambda_a=\lambda_b=\infty$, which entails the absence of persistence of market impact.

\vspace{-8pt}
\subsection{Our DRL Approach}



\begin{figure}
    \includegraphics[width=1\textwidth]{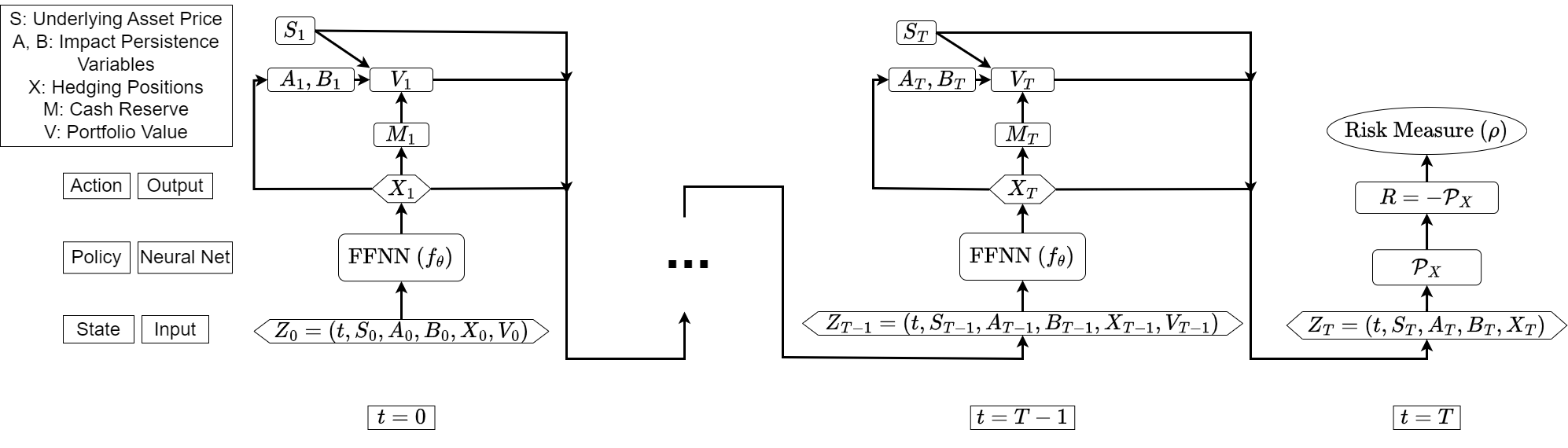}
    \caption{Neural Network Structure of Our Proposed Model.}
    \label{FIG:RNN}
\end{figure}

The goal of dynamic hedging is to find the optimal policy $X^*$ to minimize risk as defined in \cref{globhedging}. Due to the Markovian nature of the considered environment, the optimal policy is of the form $X^*_{t+1} = \tilde{X}\left(Z_t\right)$ where $\tilde{X}$ maps the state variables $Z_t \equiv \left(t,S_t,A_t,B_t,X_t,M_t\right)$ of our environment into the optimal action $X^*_{t+1}$. To approximate $\tilde{X}$, our DRL hedging approach uses a fully connected Feedforward Neural Network (FFNN) $f_\theta$ with parameters $\theta$, i.e. $\tilde{X} \approx f_\theta$. This approximation is justified by the universal approximation property of neural networks \cite{hornik1989multilayer}. Therefore, getting an approximation to the solution of \cref{globhedging} boils down to optimizing $\theta$ instead of $\tilde{X}$.
The implementation of our DRL approach is inspired by \cite{carbonneau2021equal} and a flowchart representing the structure of our DRL hedging model is given in \cref{FIG:RNN}. As shown in \cref{FIG:RNN}, the network is used with a recurrent connection for training, due to the entire time series of state variables being inputs. The input to a time-$t$ action is $Z_t$, where,
similarly to \cite{buehler2019deep}, the optimization problem is tackled through a policy gradient RL algorithm, which is closely related to stochastic gradient descent (SGD). We denote by $\hat{\rho}(R)$ the approximation of the application of the risk measure $\rho$ on the variable $R=-\mathcal{P}_X$ based on a Monte-Carlo simulation where a sample $\mathbb{B}$ of $N$ i.i.d. copies of $R$, denoted $\mathbb{B} \equiv \{R^{(1)},\ldots,R^{(N)} \}$, is generated and the distribution of $R$ is approximated by the sample's empirical distribution. The policy gradient algorithm produces a sequence of parameter sets $\{\theta_j\}^\infty_{j=0}$ that lead to progressively more refined policies characterized by $\{ f_{\theta_j} \}^\infty_{j=0}$. Such algorithm is
\begin{equation*}
\theta_{j+1} = \theta_{j} - \eta_j \nabla_\theta \widehat{\rho}(\mathbb{B}_j), \quad j=0,1,\ldots
\end{equation*}
where the learning rates sequence $\{\eta_j\}_{j \geq 1}$ contains small positive real numbers determined by the Adam optimizer \citep{kingma2014adam}, $\mathbb{B}_j$ is a minibatch containing i.i.d. sampled values of hedging losses $-\mathcal{P}_{ f_\theta(Z_{0:T-1}) }$ obtained with the current policy provided by $\theta_j$, and $\nabla_\theta$ denotes the gradient operator with respect to $\theta$. 

We implemented our approach using the PyTorch library, which explicitly computes the gradient. We normalized the parameters $S_t$ to $log\big(\frac{S_t}{K}\big)$, $V_t$ to $\frac{V_t}{V_0}$ and $t$ to $\frac{t}{T}$. Furthermore, we replaced $M_t$ with the time-$t$ portfolio value defined as: $V_t = M_t + F^b(S_t, (X_t)^+,B_t) - F^a(S_t, (-X_t)^+, A_t)$, as the model was found to converge quicker. All experiments were run on an NVIDIA RTX A4500 GPU and took around 2 hours to complete. 
\section{Experimental Results}
\label{se:NE}

To evaluate our model, we conducted several simulations adjusting various parameters and compared the behavior of our DRL-provided optimal policies with that of baselines. We performed two sets of analyses: one with no impact persistence (see \cref{se:Model}) and one with impact persistence (see \cref{se:Pinrisk}). 

\subsection{Baselines}
\label{se:QuadHedgeIssues}

To evaluate our DRL model's optimal policies $X^*$, we compared them to two common hedging strategy baselines, the Black-Scholes \cite{BSM1973} and the Leland \cite{leland1985} delta-hedges. 

The first baseline is the Black-Scholes delta hedging procedure from \cite{BSM1973}, which sets:
\begin{equation}
\label{DeltaHedge}
X_t = \Phi(d_1), \quad d_1 = \frac{\log(S_t/K) +(r+\sigma^2/2)(T-t)\delta_t}{\sqrt{\sigma^2(T-t)\delta_t}}, \quad \text{for all } t=0,\ldots,T
\end{equation}
where $\Phi$ is the standard normal cumulative distribution function. In absence of market impacts and in the continuous-time limit, this procedure is shown to completely eliminate risk, which explains its popularity. 

The second hedging strategy used as a baseline is the adaptation of delta hedging proposed by Leland \cite{leland1985}, which accounts for proportional transaction costs. The Leland strategy makes an adjustment in the volatility parameter that is used in the delta hedging formula, which consists in replacing $\sigma$ in \cref{DeltaHedge} with:
\begin{equation}
\label{Leland}
\tilde{\sigma} = \sigma \sqrt{1+\sqrt{\frac{2}{\pi}} \times \frac{k}{\sigma\sqrt{\delta_t}}}
\end{equation}
where $k$ is a proportional transaction cost rate to be adjusted in our setting. In all experiments, the parameter $k$ is set so that the revenue of buying one share of the underlying asset at the initiation date with the proportional fees $k$ is equal to that of the DRL model $F^b(0, 0, 1)$.


\subsection{Simulation Analysis: The No Impact Persistence Case}
\label{se:Model}

The first set of analyses assumes no impact persistence (i.e. $\lambda_a=\lambda_b=\infty$), which means that market transactions have only an immediate impact (i.e. trades do not affect prices at subsequent time steps $t+1,\dots, T$). Such an assumption is reasonable for larger time horizons when rebalancing is less frequent, due to the market having more time to replenish orders after a transaction.

Parameters of the geometric Brownian motion of \cref{Bsmodel} used to simulate the underlying asset prices are maximum likelihood estimates on a time series of S\&P 500 price returns extending between 1986-12-31 and 2010-04-01, which are $\mu = 0.0892$ and $\sigma = 0.1952$.
A standard call option with strike price $K=1000$ and one-year expiry with monthly time steps is hedged ($T=12$, $\delta_t=\frac{1}{12}$).
For simplicity, the risk-free rate is set to $r=0$.
The absence of impact persistence is reflected by setting $\lambda_a=\lambda_b=\infty$ and thus $A_t=B_t=0$ for all time points $t$, meaning market impacts immediately vanish after a trade. 
The following three sets of market impact parameters are considered: $\alpha=\beta=1$ (infinite liquidity); $\alpha= 1.01$, $\beta=0.99$ (high liquidity); $\alpha=1.02$, $\beta=0.98$ (low liquidity).





\subsubsection{Effect of the Hedging Portfolio Value ($V$)}
\label{se:PortfolioShadow}
\begin{figure}[h]
\centering
\subfloat[No market impact ($\alpha=\beta=1.00$)]{\includegraphics[width=0.5\textwidth]{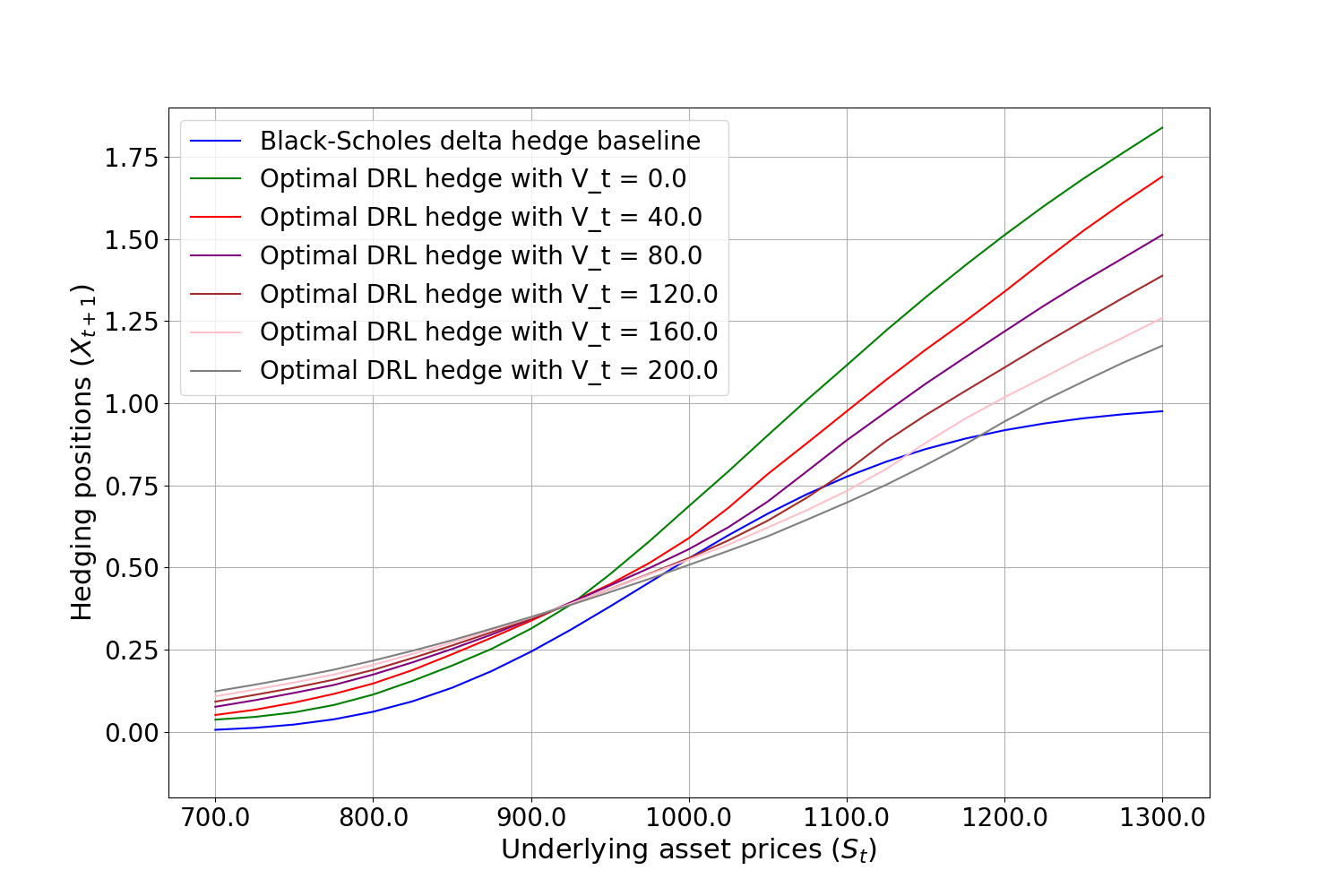}}
\subfloat[With market impact  ($\alpha=1.01$, $\beta=0.99$)] {\includegraphics[width=0.5\textwidth]{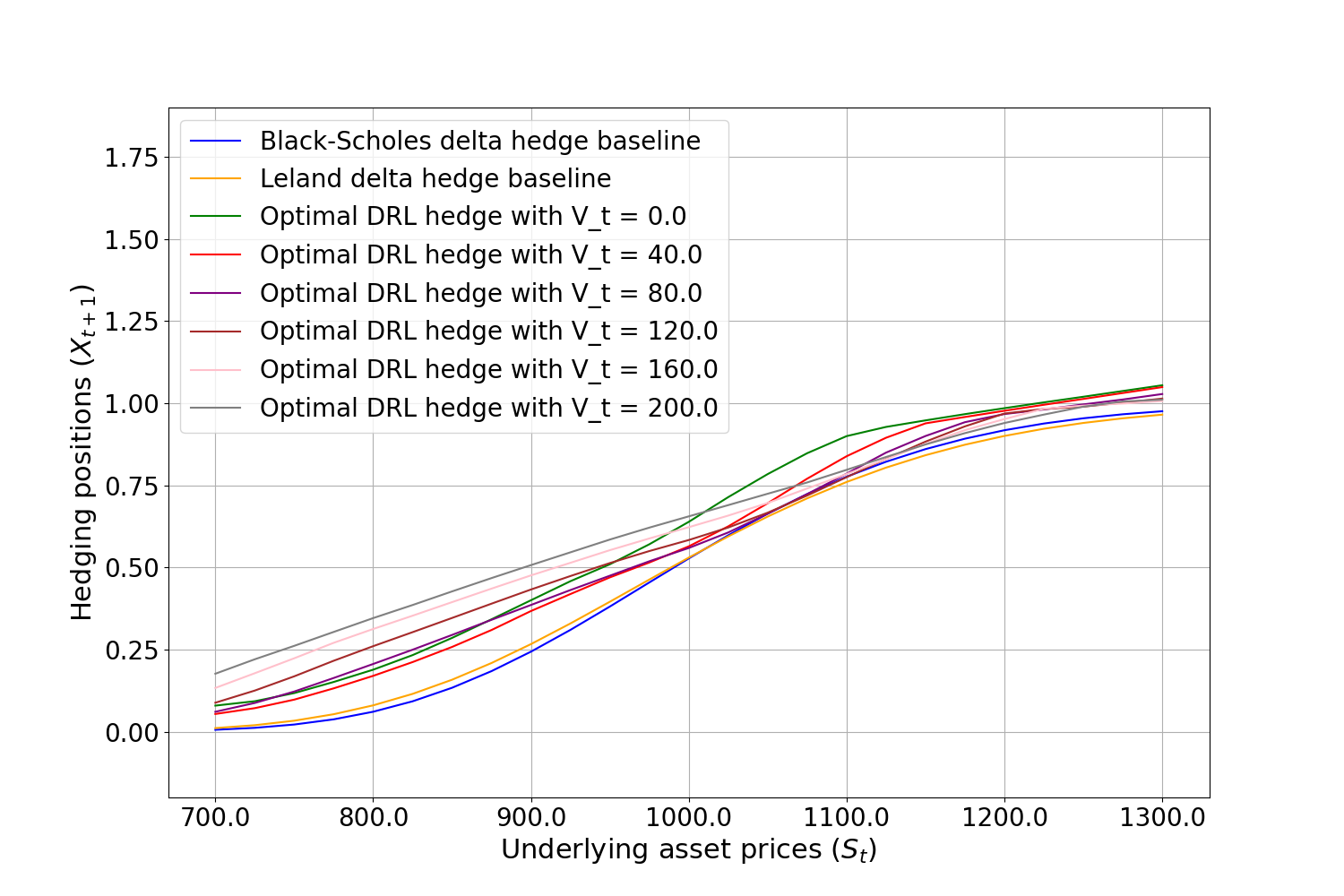}}
        \caption{\label{FIG:HedgingPolicy1} Evaluation of hedging positions $X_{t+1}$, at six months-to-maturity ($t=6, T=12, \delta_t=\frac{1}{12}$) with monthly rebalancing for a call option with strike price $K=1000$. 
        }
\end{figure}

We first analysed the optimal hedging position $X_{t+1}$ at $t=6$  (i.e. halfway through the option lifetime) as a function of the underlying asset price $S_t$ for various hedging portfolio values $V_t$.
The previous period hedging position is assumed to be $X_{t}=0.5$.
Such hedging portfolio composition is reported in \cref{FIG:HedgingPolicy1} for the semi-quadratic penalty risk measure $\rho^{semi}$ (see \cref{se:marketdyn}). The left panel reports positions in absence of market impact ($\alpha=\beta=1$), whereas optimal positions under low market impact ($\alpha=1.01$, $\beta=0.99$) are provided in the right panel. The Black-Scholes delta hedging baseline strategy (in blue) is displayed on both panels and the Leland delta hedging strategy (in orange) is displayed only on the right panel because the two baselines are equivalent when omitting market impact. 




\vspace{-10pt}

As \cref{FIG:HedgingPolicy1} shows, a key difference between the DRL-based optimal hedging positions and those of the baselines is that our optimal DRL strategy depends on the hedging portfolio value at time $t$ $(V_t)$ whereas both delta hedging strategies ignore $V_t$ in their decisions. Indeed, our hedging method adjusts its position based on how much hedging gains/losses it has already incurred up to the current time step $t$, needing to act more conservatively if a higher hedging loss is likely to be incurred at the expiry $T$. The ability of our hedging method to look backwards and correct for previous hedging errors is a favourable feature of our approach that improves over the baselines.
The impact of an increase of $V_t$ changes depending on the underlying asset price level $S_t$. For instance, in absence of market impacts as depicted in Panel (a), our DRL model's optimal position $X_t$ is increasing (resp. decreasing) in $V_t$ for low (resp. high) values of $S_t$.
In Panel (a), when $S_t$ is very low, optimal positions that are increasing in terms of $V_t$ occur because larger positions are desirable if the underlying price increases subsequently, as it would lead to larger gains better offsetting the increase in the option value. Conversely, larger positions would be less desirable if the price declines in future periods due to generating future losses. However, the larger $V_t$ is, the less problematic such losses are as a larger reserve of cash makes hedging shortfalls at maturity less likely. As such, a higher $V_t$ allows to better protect against the most adverse cases involving a future underlying asset price increase (and thus an appreciation of the option value) by tilting the hedging position upward, as it would make future moderate possible losses due to a market decline more easily absorbed. Indeed, the risk caused by an increase in the option value can be much more severe due to the option price value not having an upper bound, whereas option values are bounded below by zero which limits risk stemming from an asset price decline.

\subsubsection{Comparison of Hedging Position Sequences ($X$)}
\label{se:PortfolioPaths}
\begin{figure}[h]
\centering
\subfloat[Underlying asset price sequence $S=\{S_t\}^T_{t=0}$]{\includegraphics[width=0.5\textwidth]{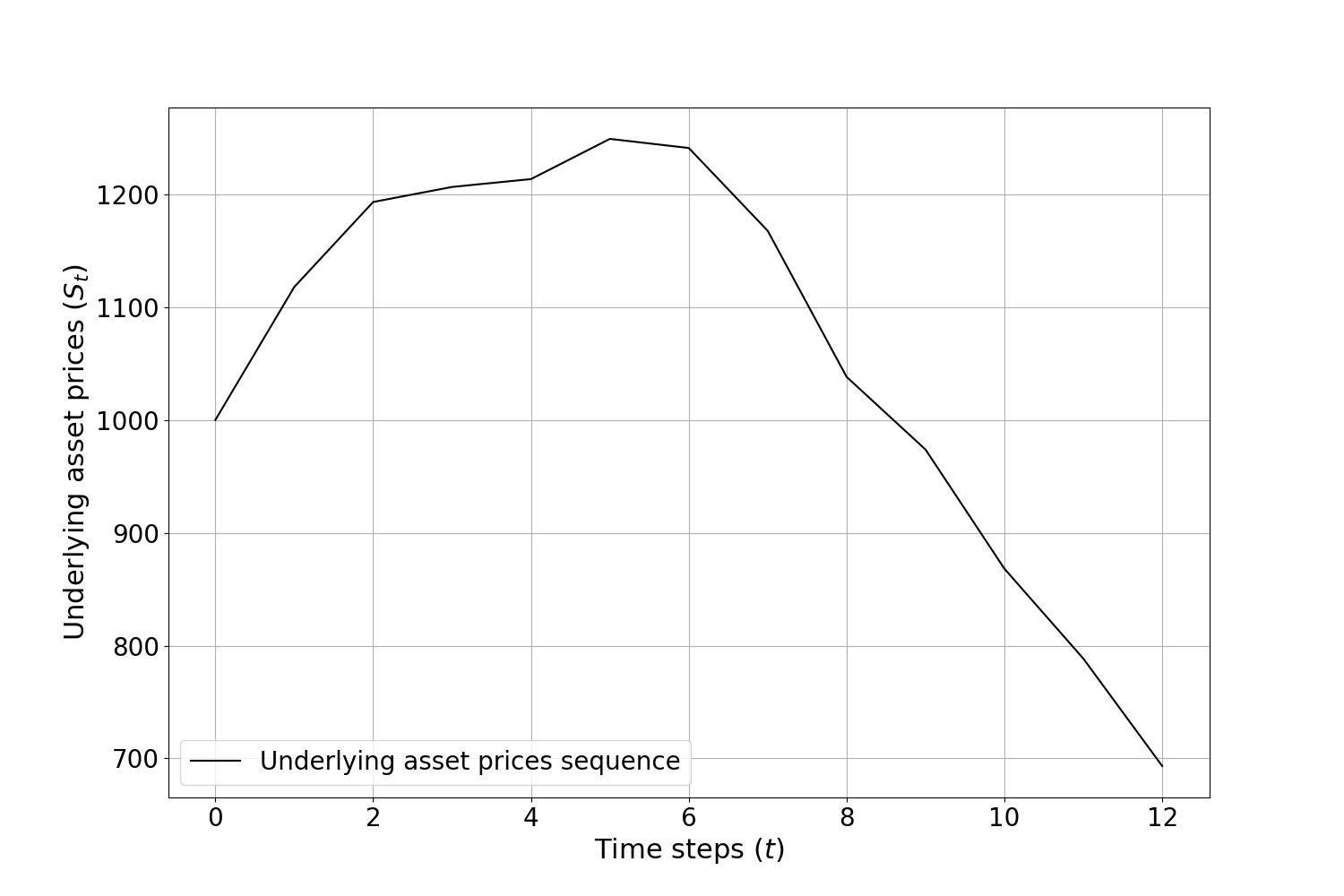}}
\subfloat[Hedging position sequence $X=\{X_{t+1}\}^{T-1}_{t=0}$ (underlying asset shares)] {\includegraphics[width=0.5\textwidth]{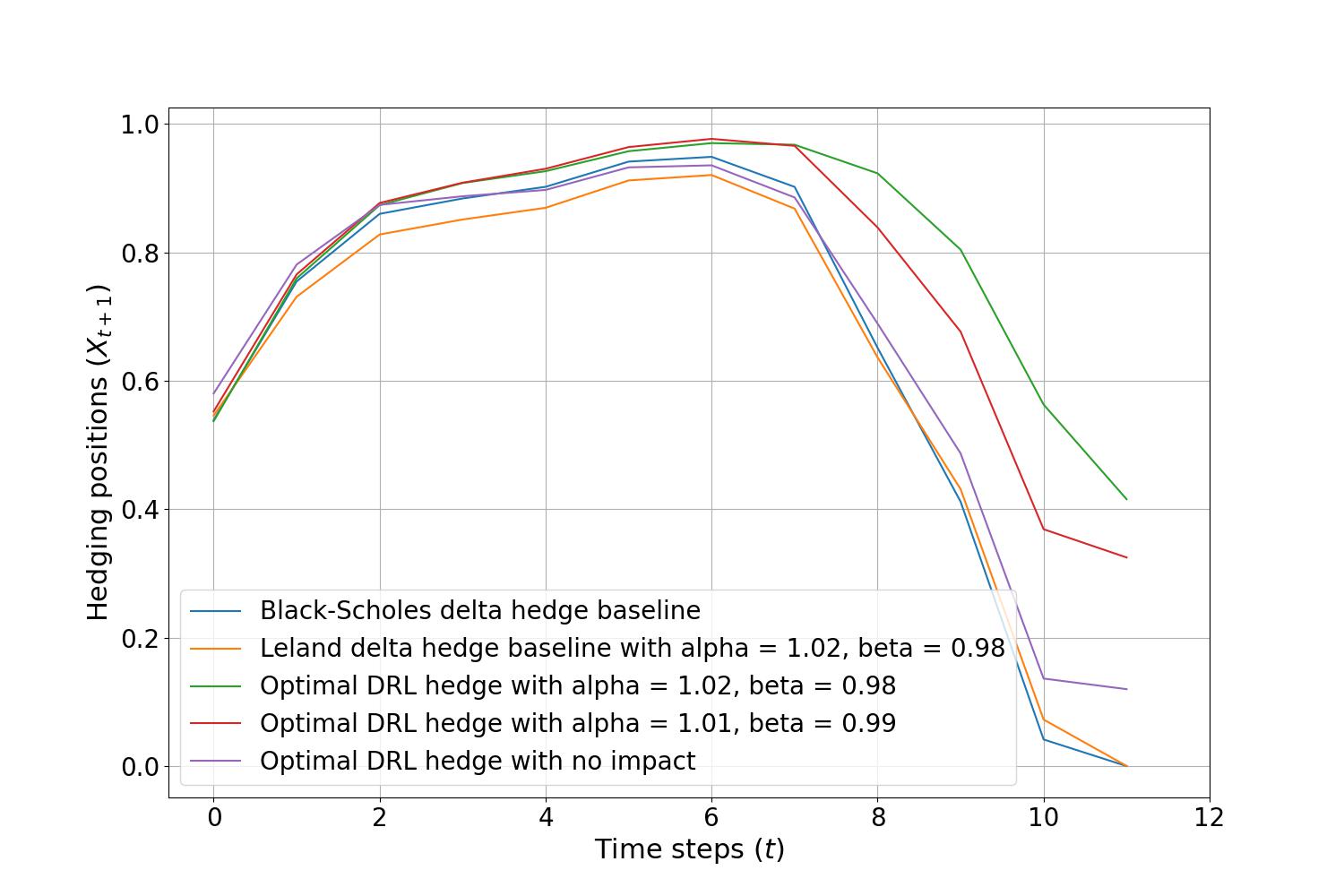}} 
        \caption{\label{FIG:SimulPathHedgeYearly} Evaluation of the hedging position sequence $\{X_{t+1}\}^{T-1}_{t=0}$ for a simulated underlying asset price sequence $S=\{S_t\}^T_{t=0}$ with monthly rebalancing for a one-year-to-maturity ($T=12,\delta_t=\frac{1}{12}$) call option with strike price $K=1000$. 
        }
\end{figure}

The second analysis we conducted consists in studying the evolution of the hedging position (i.e. the sequence of actions $X=\{X_{t+1}\}^{T-1}_{t=0}$) over various simulated paths of the underlying asset price. 
To evaluate this, we randomly generated an underlying asset path based on the geometric Brownian motion of \cref{Bsmodel}.
\cref{FIG:SimulPathHedgeYearly} reports the hedging positions for different levels of market impact. Panel (a) shows the sequence of simulated underlying asset prices $S=\{S_t\}^T_{t=0}$. Panel (b) shows the hedging positions sequence $X=\{X_{t+1}\}^{T-1}_{t=0}$ (the DRL model's optimal hedging positions and the baselines hedging positions), for various market impact parameters $\alpha,\beta$, where $X_0=0$.
For all models, the hedging positions start by behaving similarly and then slowly diverge when getting closer to the expiry $T$. 
\vspace{-10pt}
As the level of liquidity decreases (i.e. $\alpha$ and $\beta$ becoming further away from $1$), our DRL model's optimal hedging positions (purple, red and green lines) stray further away from the delta hedging baseline positions (blue and orange lines) by performing smaller adjustments when rebalancing due to their higher cost. As expected, taking market impact into account while making hedging decisions dampens the turnover of the strategy. Wider differences in the hedging positions between levels of market impact at later stages can be explained by 1) the fact that the portfolio value $V_t$ differs more widely across strategies when nearing the expiry due to having spent more periods with different positions, and 2) a time-to-maturity effect. Indeed, properly adjusting hedging positions when the time-to-maturity is low becomes less important, as the magnitude of market risk faced by the agent until maturity shrinks when less periods remain before expiry. As such, in the presence of market impacts, late-period positions are further from these without market impacts (i.e. ideal positions) since the reduction in market risk would not be sufficient to justify incurring the market impact costs.

\subsubsection{Effect of the Drift ($\mu$)}
\label{se:ttmYear}
\begin{figure}[h]
\centering
\subfloat[Hedging position sequence $X=\{X_{t+1}\}^{T-1}_{t=0}$\\ for $\mu=0.0892$] {\includegraphics[width=0.5\textwidth]{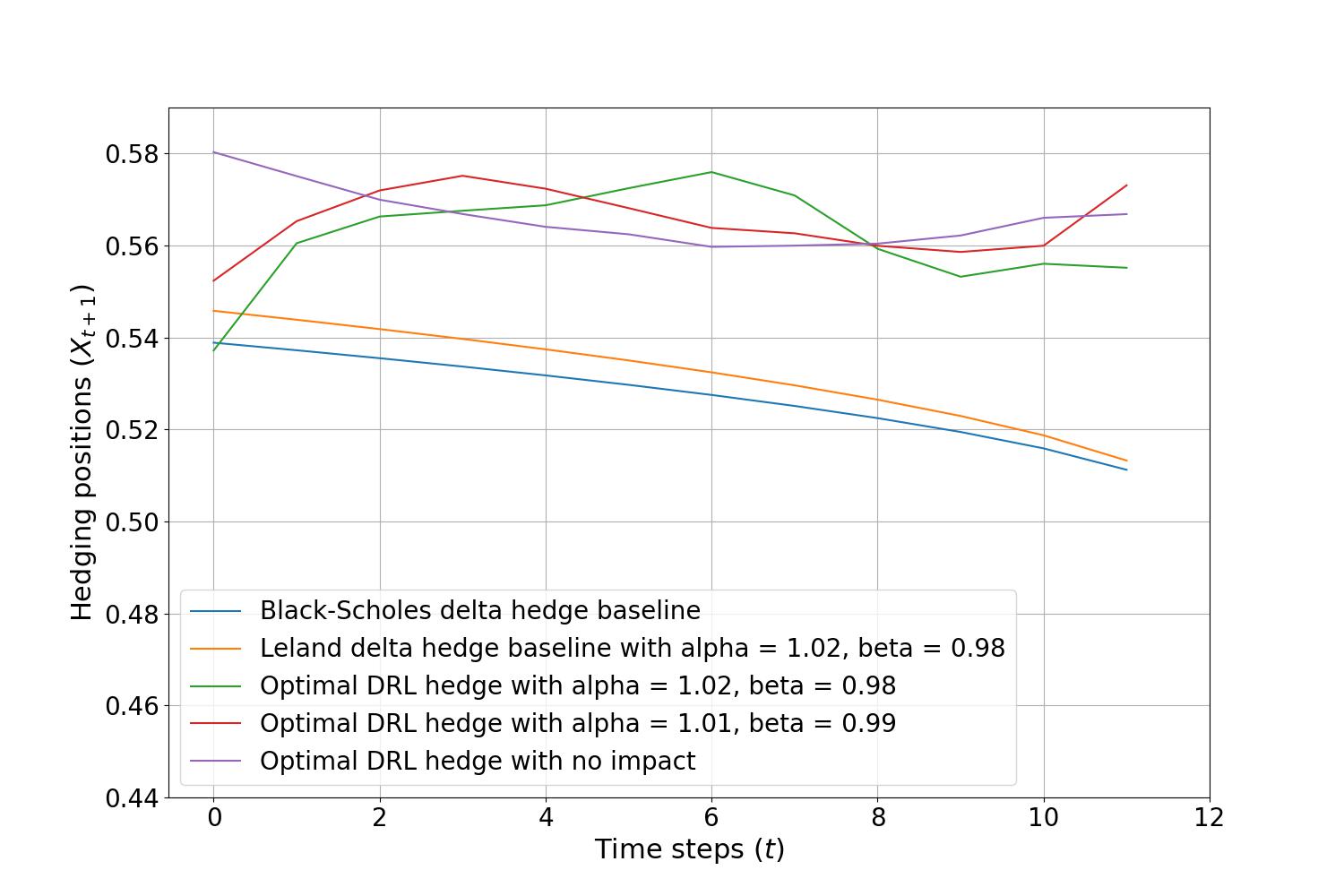}} 
\subfloat[Hedging position sequence $X=\{X_{t+1}\}^{T-1}_{t=0}$ for $\mu=0.0$] {\includegraphics[width=0.5\textwidth]{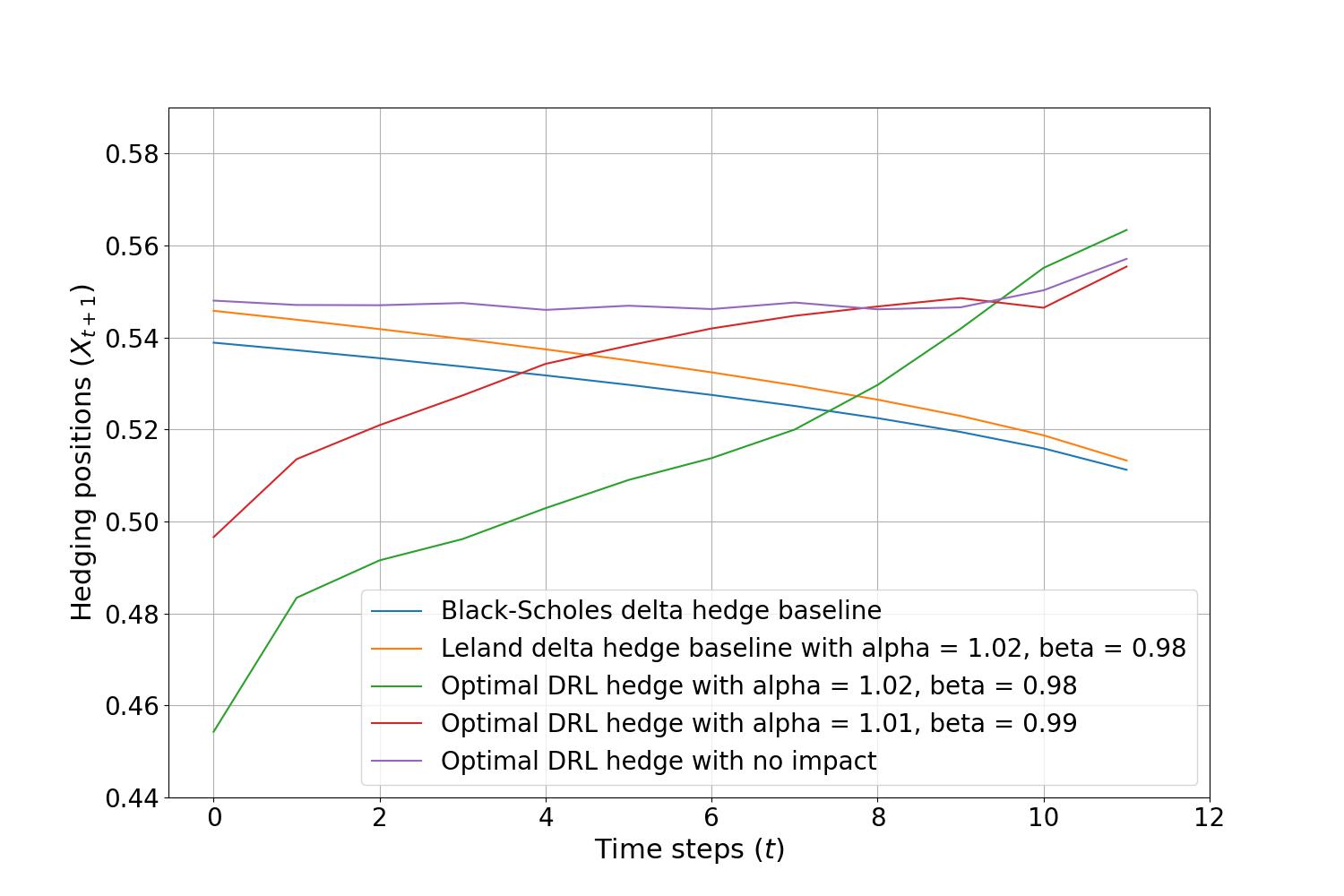}}
        \caption{\label{FIG:SimulPathHedge1Strike} Evaluation of the hedging position sequence $\{X_{t+1}\}^{T-1}_{t=0}$ where the underlying asset price sequence remains at the strike price with monthly rebalancing for a one-year to maturity ($T=12, \delta_t=\frac{1}{12}$) call option with strike price $K=1000$. 
        }
\end{figure}

A well-known property of delta hedging methods is that hedging positions do not consider the drift parameter (the annualized expected return parameter $\mu$ from \cref{Bsmodel}). However, our DRL model's optimal hedging positions do implicitly take into account the drift. To evaluate its effect, we considered a second fictitious sequence of underlying asset prices. The price on such a sequence remains constant until maturity, i.e. $S_t=1000$, $t=0,\ldots,T$. The same market dynamics as in \cref{se:PortfolioPaths} were used. 
The optimal hedging position sequence of our DRL model for different levels of market impact, along with delta hedging positions, are shown in Panel (a) of \cref{FIG:SimulPathHedge1Strike}. Furthermore, to assess the impact of the drift parameter $\mu$, a parameter that is difficult to estimate with precision in practice, Panel (b) of \cref{FIG:SimulPathHedge1Strike} shows the DRL model's optimal hedging position sequence provided by a model that was trained under sequences generated with $\mu=r=0$ instead of $\mu=0.0892$, along with the corresponding baseline strategies (the orange and blue curves are identical in both panels).



\vspace{-10pt}

When the network is trained on underlying asset sequences $S=\{S_t\}^T_{t=0}$ generated with the drift $\mu=0.0892$, our DRL model's optimal hedging positions for all three levels of market impact $\alpha$ and $\beta$ are larger than these of both delta hedge strategies at most time steps $t$. Our DRL model's hedging positions from Panel (b) with null drift ($\mu=0$) are significantly smaller than positions with the positive drift from Panel (a), especially during initial time steps. Increasing the drift therefore leads to our model buying more shares of the underlying asset, which leads to higher profitability of trades, thereby reducing risk.

In the first few initial time steps, the differences between our DRL model's hedging positions for different levels of market impact are larger for earlier time steps $t$. This is more apparent for the case of null drift ($\mu=0$) from Panel (b). This happens due to the correction of inadequate hedging positions being gradually deferred through time to avoid incurring costs stemming from market impacts, with higher impacts leading to lengthier deferral. For the case of the positive drift ($\mu=0.0892$), such phenomenon is less noticeable across the levels of market impact as increased profitability associated with stock purchases compensate for the market impacts incurred.

In summary, the drift can have a significant impact on our DRL model's optimal hedging positions by increasing underlying asset positions as the drift increases. In addition, market impact is seen to partially defer the correction of inadequate hedging positions, with temporary increased market risk sometimes being considered more tolerable than high market costs that would be generated by immediately setting up the desired position.


\subsection{Simulation Analysis: Impact Persistence in the Case of Pin Risk}
\label{se:Pinrisk}

We now turn to the case of a market environment which includes impact persistence (i.e. $\lambda_a < \infty$ and $\lambda_b < \infty$). Recall that impact persistence refers to market impacts of a purchase or a sale persisting over time. We will specifically analyse the case of pin risk which refers to a situation where the underlying stock price $S_t$ is very close to the option strike price $K$ and the time-to-maturity $T-t$ is very small. Properly hedging in this situation is very hard; the gamma (second order partial derivative of the option price with respect to the underlying asset price) of the option explodes, as the delta ($X_t$ from \cref{DeltaHedge}) of the option quickly oscillates from close to $0$ to close to $1$ (and vice-versa) depending on whether the underlying stock price is below or above the option strike price ($S_t<K$ or $S_t>K$). Such instability is very problematic in the presence of market impact due to very large costs that would need to be incurred if the hedging portfolio was rebalanced abruptly as prescribed by delta hedging. We therefore turn to our DRL methodology to analyse how to improve on hedging procedures in the very challenging situation of pin risk. 
When nearing maturity ($t$ close to $T$) and facing smaller time scales (small $\delta_t$), the assumption of not having impact persistence might become unrealistic, and incorporating impact persistence in the limit order book model might be key to increase realism, which we do in this section.

The case of hedging an 8-hour-to-maturity call option with hourly rebalancing steps ($T=8, \delta_t=\frac{1}{8\times252}$, with $252$ being the number of business days in a year) is considered.
The initial underlying stock price is $S_0=K=1000$. Drift and volatility parameters are still $\mu=0.0892$ and $\sigma=0.1952$. Market liquidity parameters are set to $\alpha=1.001 \text{ and } \beta=0.999$. Various degrees of impact persistence are considered: $\lambda_a=\lambda_b$ are being set to $\infty$ (impacts have no persistence), $-\ln(0.5)$ (impact decaying by $50\%$ at each time step), or $0$ (impacts are permanent). Again, we set $X_0=0$.

\begin{figure}[htb]
\centering
\subfloat[Underlying asset price sequence $S=\{S_t\}^T_{t=0}$]{\includegraphics[width=0.5\textwidth]{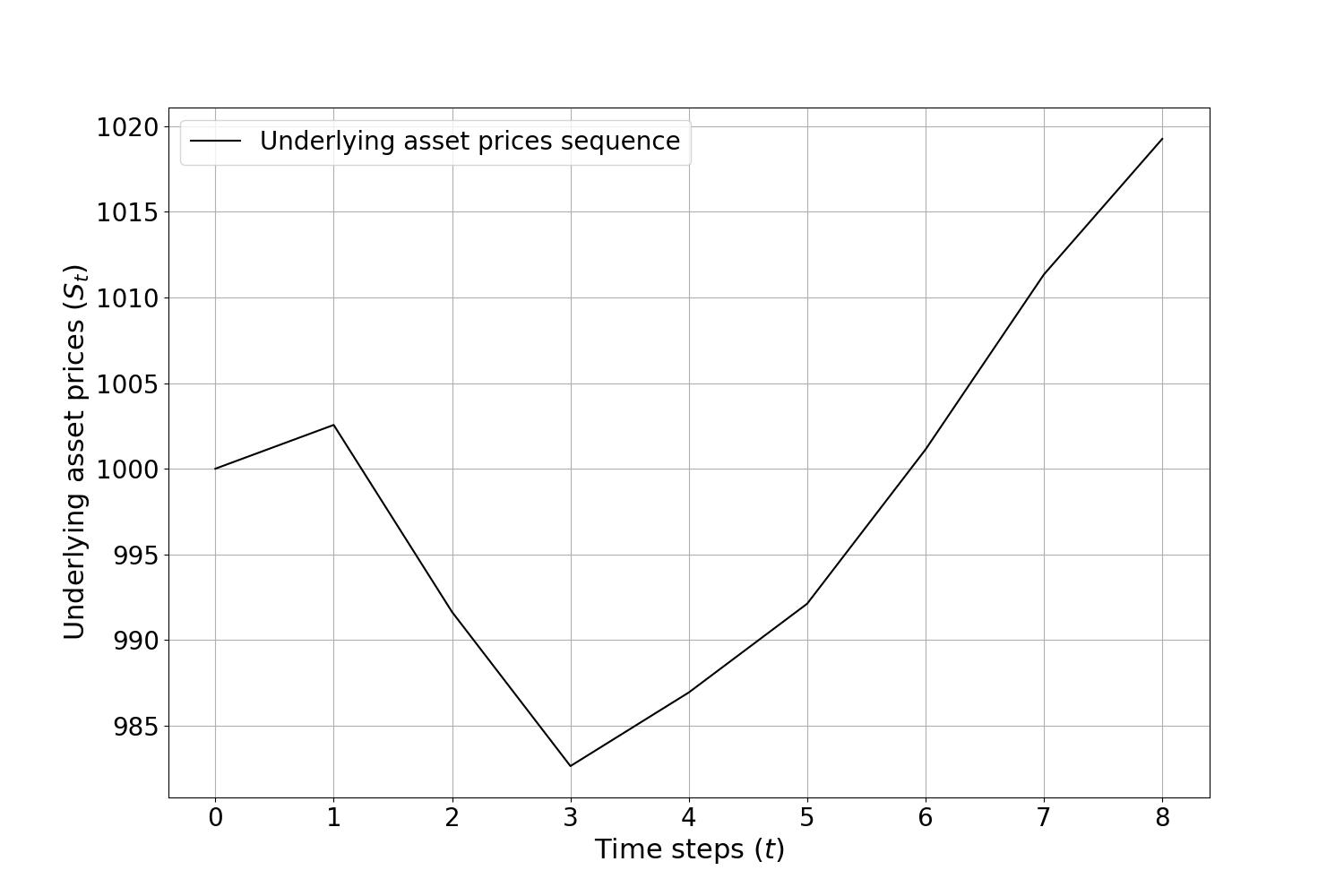}}
\subfloat[Hedging position sequence $X=\{X_{t+1}\}^{T-1}_{t=0}$] {\includegraphics[width=0.5\textwidth]{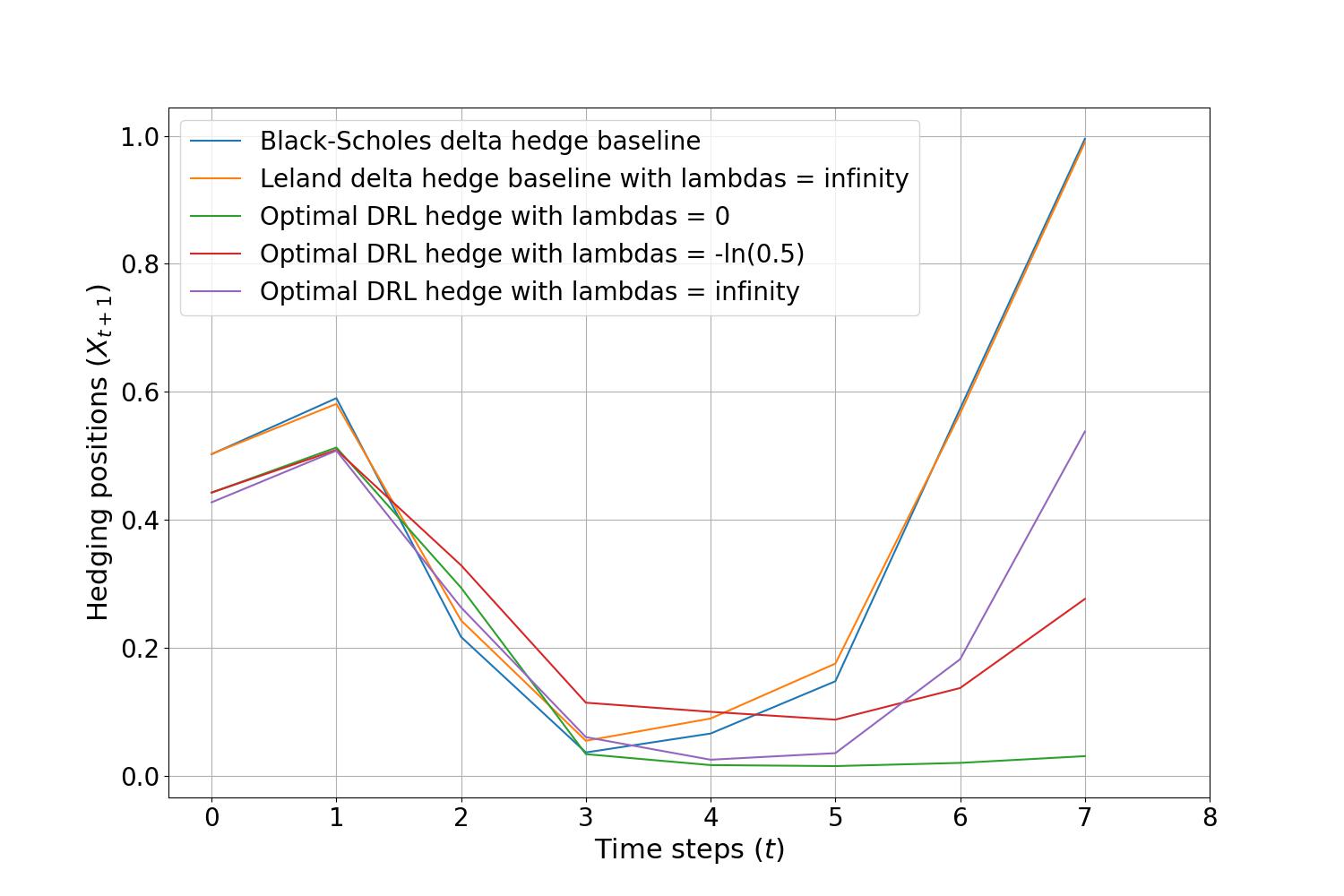}} 
        \caption{\label{FIG:SimulPathHedge1PinRiskPersistence} Evaluation of hedging position sequence $\{X_{t+1}\}^{T-1}_{t=0}$ for a simulated underlying asset price sequence $S=\{S_t\}^T_{t=0}$ with hourly rebalancing for an 8 hour to maturity ($T=8,\delta_t=\frac{1}{8\times252}$) call option with strike price $K=1000$. Market liquidity parameters $\alpha=1.001$ and $\beta=0.999$ are used. 
        }
\end{figure}

\cref{FIG:SimulPathHedge1PinRiskPersistence} reports the evolution of hedging positions for a simulated sequence of underlying asset prices, considering different levels of impact persistence. Panel (a) shows the sequence of simulated underlying asset prices $S=\{S_t\}^T_{t=0}$.
Panel (b) shows the hedging positions sequence $X=\{X_{t+1}\}^{T-1}_{t=0}$ (these of the DRL model and of the baselines), for various impact persistence parameters $\lambda_a \text{ and } \lambda_b$. 
The purple line corresponds to the DRL model without market impact persistence where impacts decay immediately ($\lambda_a=\lambda_b=\infty$). The red line corresponds to the case where market impacts  decays by $50\%$ at each time step $t$, and the green line represents the case where there is permanent market impact which does not decay over time ($\lambda_a=\lambda_b=0$).
\vspace{-10pt}
As we increase the market impact persistence, the DRL model's hedging positions for time steps near maturity exhibit smaller fluctuations. Indeed, agents working in a high persistence environment cannot overcome the highly persistent market impacts caused by hedging rebalancing actions taken at previous time steps; the model therefore suggests refraining from buying the underlying asset as much as possible to avoid incurring high costs. The effect of impacts is more severe near maturity since they had time to accumulate.
However, the relation between the size of the hedging position $X_t$ and the impact persistence parameters $\lambda_a,\lambda_b$ is not monotonic; between time steps $t=1$ and $t=5$, the DRL model with intermediate persistence, namely $\lambda_a=\lambda_b=-\ln(0.5)$ (red line), takes larger hedging positions than DRL strategies with both null or permanent persistence. This indicates a complicated relationship between the hedging positions $X_t$ and the levels of impact persistence ($\lambda_a,\lambda_b$), which highlights the necessity of relying on complex optimization schemes such as DRL to uncover the optimal policy. Furthermore, parameters driving market impacts can interact and push optimal positions in different directions, with the net impact being unclear \textit{à priori}. For instance, while large values of illiquidity parameters $\alpha,\beta$ incentivize partially delaying the correction of inadequate hedging positions as discussed earlier, large impact persistence parameters $\lambda_a,\lambda_b$ offset this effect by making the delaying approach less effective, with market impacts still being felt in subsequent time steps.
\section{Conclusion and Further Work}
\label{se:conclusion}

In this paper, we investigated the use of Deep Reinforcement Learning for the problem of optimizing the dynamic hedging of an option in the presence of market impacts. Our Deep Reinforcement Learning (DRL) approach is inspired by the deep hedging approach proposed for instance in \cite{buehler2019deep} and \cite{carbonneau2021equal} to solve the associated sequential decision problem; a policy gradient algorithm is considered to train the FFNN mapping market environment state variables to hedging actions. However, we consider an augmented state space by embedding the model of \cite{simard2014general} to characterize market impact, which is a crucial and necessary addition to achieve proper performance in markets where liquidity issues are prevalent. To the best of our knowledge, our work is the first to integrate market impact and limit order book models in a deep hedging setting. 

A key contribution of our work is 
to analyse the optimal policy obtained in the simulated market environment so as to gain insight on how the various state variables impact optimal decisions. 
We believe such a contribution is very important, as multiple studies about optimal hedging analyse the magnitude of out-performance gained from using an optimal strategy over a baseline approach, without necessarily scrutinizing the optimal policy itself or explaining how such higher performance is attained. Through several sensitivity analyses and simulations, the following interesting findings were obtained:
\begin{itemize}
    \item The hedging portfolio value $V_t$ and the underlying asset drift (i.e. expected return) parameter $\mu$ materially impact optimal positions. Such features are not considered by the delta hedging baseline approach.
    \item The deep hedging method learns to what extent the size of portfolio rebalancing operations must be dampened or delayed to avoid paying excessive market-impact-related costs while still providing adequate market risk protection.
    \item Differences between optimal hedging positions and that of conventional delta hedging baselines result from complex interactions between multiple variables, including time-to-maturity $T-t$, the impact magnitude parameters $\alpha,\beta$ and parameters $\lambda_a,\lambda_b$ driving impact persistence.
\end{itemize}
The interrelated (and sometimes competing) effects of the multiple parameters and state variables driving hedging decisions highlight the importance of using sophisticated approaches such as DRL to determine optimal hedging policies, as proposing simple rule-of-thumb adjustments to apply to delta hedging strategies is most likely insufficient to reach near-optimal performance in the context of illiquid markets.

As further work, we could consider modeling the entire limit order book in a data-driven fashion instead of relying on simplifying dynamics \cref{dynamicsF}-\cref{dynamicsG} for market impacts. Such enhancement would require testing whether the FFNN-based policy gradient algorithms would be sufficient to treat high-dimensional state spaces required to deal with such framework, or if a different Neural Network architecture is needed. 
\printbibliography[heading=subbibintoc]

@article{almgren2001optimal,
  title={Optimal execution of portfolio transactions},
  author={Almgren, Robert and Chriss, Neil},
  journal={Journal of Risk},
  volume={3},
  pages={5--40},
  year={2001}
}

@ARTICLE{Almgren/Li_2016,
TITLE   = {Option hedging with smooth market impact},
AUTHOR  = {R. Almgren and T. M. Li},
JOURNAL = {Market Structure and Liquidity},
VOLUME  = {2},
NUMBER  = {1},
YEAR    = {2016},
PAGES   = {},
}

@ARTICLE{Bank/Baum_2004,
TITLE =  {Hedging and portfolio optimization in illiquid financial markets with a large trader},
AUTHOR =  {Bank, P. and Baum, D.},
JOURNAL =  {Mathematical Finance},
VOLUME = {14},
YEAR = {2004},
PAGES = {1--18}}

@article{BSM1973,
    author = {Fischer Black and Myron Scholes},
    year = {1973},
    title = {The pricing of options and corporate liabilities},
    journal = {The Journal of Political Economy},
    volume = {81},
	number = {3},
	pages = {637--654}
}

@article{buehler2019deep,
  title={Deep hedging},
  author={Buehler, Hans and Gonon, Lukas and Teichmann, Josef and Wood, Ben},
  journal={Quantitative Finance},
  volume={19},
  number={8},
  pages={1271--1291},
  year={2019},
  publisher={Taylor \& Francis}
}

@ARTICLE{Calvez/Cliff_2018 ,
TITLE   = {Deep learning can replicate adaptive traders in a limit-order-book finanical market},
AUTHOR  = {A. le Calvez and D. Cliff},
JOURNAL = {Proceedings of IEEE Symposium on Computational Intelligence for Financial Engineering},
VOLUME  = {18},
NUMBER  = {21},
YEAR    = {2018},
PAGES   = {}}

@article{carbonneau2021deep,
  title={Deep equal risk pricing of financial derivatives with non-translation invariant risk measures},
  author={Carbonneau, Alexandre and Godin, Fr{\'e}d{\'e}ric},
  journal={Risks},
  volume={11},
  number={8},
  pages={140},
  year={2023},
  publisher={MDPI}
}

@article{carbonneau2021equal,
  title={Equal risk pricing of derivatives with deep hedging},
  author={Carbonneau, Alexandre and Godin, Fr{\'e}d{\'e}ric},
  journal={Quantitative Finance},
  volume={21},
  number={4},
  pages={593--608},
  year={2021},
  publisher={Taylor \& Francis}
}

@article{cartea2015optimal,
  title={Optimal execution with limit and market orders},
  author={Cartea, Alvaro and Jaimungal, Sebastian},
  journal={Quantitative Finance},
  volume={15},
  number={8},
  pages={1279--1291},
  year={2015},
  publisher={Taylor \& Francis}
}

@ARTICLE{Cetin/Jarrow/Protter_2004,
TITLE   = {Liquidity risk and arbitrage pricing theory},
AUTHOR  = {Cetin, U. and Jarrow, R.A. and Protter, P.},
JOURNAL = {Finance and Stochastics},
VOLUME  = {8},
YEAR    = {2004},
PAGES   = {311--341}}

@ARTICLE{Cetin/Soner/Touzi_2010 ,
TITLE   = {Option hedging for small investors under liquidity costs},
AUTHOR  = {Cetin, U. and Soner, H.M. and Touzi, N.},
JOURNAL = {Finance and Stochastics},
VOLUME  = {14},
NUMBER  = {},
YEAR    = {2010},
PAGES   = {317--341}}

@ARTICLE{Ellersgaard/Tegner_2017 ,
TITLE   = {Optimal hedge tracking portfolios in a limit order book},
AUTHOR  = {S. Ellersgaard and M. Tegnér},
JOURNAL = {Market Microstructure and Liquidity},
VOLUME  = {3},
NUMBER  = {2},
YEAR    = {2017},
PAGES   = {},
}

@article{godin2016minimizing,
  title={Minimizing {CV}a{R} in global dynamic hedging with transaction costs},
  author={Godin, Fr{\'e}d{\'e}ric},
  journal={Quantitative Finance},
  volume={16},
  number={3},
  pages={461--475},
  year={2016},
  publisher={Taylor \& Francis}
}

@ARTICLE{Guasoni/Weber_2020,
TITLE   = {Nonlinear price impact and portfolio choice},
AUTHOR  = {P. Guasoni and M. H. Weber},
JOURNAL = {Mathematical Finance},
VOLUME  = {30},
NUMBER  = {},
YEAR    = {2020},
PAGES   = {341--376},
}

@article{gueant2017option,
  title={Option pricing and hedging with execution costs and market impact},
  author={Gu{\'e}ant, Olivier and Pu, Jiang},
  journal={Mathematical Finance},
  volume={27},
  number={3},
  pages={803--831},
  year={2017},
  publisher={Wiley Online Library}
}

@book{hull1993options,
  title={Options, futures, and other derivative securities},
  author={Hull, John},
  volume={7},
  year={1993},
  publisher={Prentice Hall Englewood Cliffs, NJ}
}

@article{cao2023gamma,
  title={Gamma and vega hedging using deep distributional reinforcement learning},
  author={Cao, Jay and Chen, Jacky and Farghadani, Soroush and Hull, John and Poulos, Zissis and Wang, Zeyu and Yuan, Jun},
  journal={Frontiers in Artificial Intelligence},
  volume={6},
  pages={1129370},
  year={2023},
  publisher={Frontiers}
}

@article{wu2023robust,
  title={Robust Risk-Aware Option Hedging},
  author={Wu, David and Jaimungal, Sebastian},
  journal={arXiv preprint arXiv:2303.15216},
  year={2023}
}

@article{hodges1989optimal,
  title={Optimal replication of contingent claims under transaction costs},
  author={Hodges, Stewart and Neuberger, Anthony},
  journal={Review Futures Market},
  volume={8},
  pages={222--239},
  year={1989}
}

@article{hornik1989multilayer,
  title={Multilayer feedforward networks are universal approximators},
  author={Hornik, Kurt and Stinchcombe, Maxwell and White, Halbert},
  journal={Neural Networks},
  volume={2},
  number={5},
  pages={359--366},
  year={1989},
  publisher={Elsevier}
}

@article{kingma2014adam,
  title={Adam: A method for stochastic optimization},
  author={Kingma, Diederik P and Ba, Jimmy},
  journal={arXiv preprint arXiv:1412.6980},
  year={2014}
}

@article{leland1985,
  title={Option pricing and replication with transaction costs},
  author={Leland, Hayne E.},
  journal={the Journal of Finance},
  volume={40},
  number={5},
  pages={1283--1301},
  year={1985},
  publisher={Wiley for the American Finance Association}
}

@article{ning2021double,
  title={Double deep {Q}-learning for optimal execution},
  author={Ning, Brian and Lin, Franco Ho Ting and Jaimungal, Sebastian},
  journal={Applied Mathematical Finance},
  volume={28},
  number={4},
  pages={361--380},
  year={2021},
  publisher={Taylor \& Francis}
}

@ARTICLE{Rogers/Singh_2010,
TITLE   = {The cost of illiquidity and its effects on hedging},
AUTHOR  = {Rogers, L.C.G. and Singh, S.},
JOURNAL = {Mathematical Finance},
VOLUME  = {20},
NUMBER  = {4},
YEAR    = {2010},
PAGES   = {597--515}}

@article{schweizer1995variance,
  title={Variance-optimal hedging in discrete time},
  author={Schweizer, Martin},
  journal={Mathematics of Operations Research},
  volume={20},
  number={1},
  pages={1--32},
  year={1995},
  publisher={INFORMS}
}

@article{simard2014general,
  title={General model for limit order books and market orders},
  author={Simard, Clarence},
  journal={Available at SSRN 2435198},
  year={2014}
}

\end{document}